\documentclass[aps,prl,twocolumn,superscriptaddress,nofootinbib,preprintnumbers,amsmath,amssymb]{revtex4-2}

\usepackage{amsmath,amssymb,booktabs}
\usepackage{dsfont}
\usepackage{graphicx}
\usepackage[dvipsnames]{xcolor}
\usepackage{tikz}
\usepackage[normalem]{ulem}
\AtBeginDocument{\let\origcite\cite\let\origsout\sout\renewcommand{\sout}[1]{{\renewcommand{\cite}[1]{\mbox{\origcite{##1}}}\origsout{#1}}}}
\usepackage[colorlinks=true,urlcolor=blue,linkcolor=blue,citecolor=blue]{hyperref}

\newcommand{\rep}[1]{\mathbf{#1}}
\newcommand{\Tr}{\operatorname{Tr}}
\newcommand{\vev}[1]{\langle #1\rangle}

\begin{document}

\preprint{CERN-TH-2026-301}
\title{A New GUT}

\author{Prateek Agrawal}
\affiliation{Department of Physics, University of California, Santa Barbara, CA 93106, USA}
\author{Michael Nee}
\affiliation{Department of Physics, Harvard University, Cambridge, MA 02138, USA}
\author{Mario Reig}
\affiliation{Theoretical Physics Department, CERN, 1211 Geneva 23, Switzerland}

\begin{abstract}
We present a new class of 4-dimensional Grand Unified Theories where the gauge sectors of the standard model have  a novel embedding in the unified group that does not factor through an $SU(5)$ subgroup. The smallest example has gauge algebra $\frak{su}(12)$ and an anomaly-free spectrum that produces the three standard model generations with no additional chiral matter. The embedding predicts the axion-photon anomaly ratio $E/N=2/3$ and the weak mixing angle $\sin^2\theta_W=1/2$ at the unification scale. Gauge-mediated proton decay proceeds through flipped-$SU(5)$-like operators and the minimally charged smooth monopole carries two Dirac charge units. The global gauge group can be $SU(12)$ or $SU(12)/\mathbb{Z}_2$, with the former allowing color singlets of half-integer electric charge. We develop a GUT bootstrap to constrain $SU(n)$ embeddings with three generations and no additional chiral matter. For unified fermions in up to two-index representations, we prove that all new embeddings arise from the $SU(8m+4)$ family, including flavor unified models for $m \geq 2$.
\end{abstract}

\maketitle

Grand Unification offers a compelling origin story for the Standard Model (SM). Unification explains why the SM gauge couplings are all of similar order and the pattern of charges of SM fermions, in particular the quantization of electric charge in integers~\cite{Georgi:1974sy,Pati:1974yy,Georgi:1974yf,Langacker:1980js}.

The classic example of a grand unified theory (GUT) is the $SU(5)$ GUT~\cite{Georgi:1974sy}. All the simple GUTs\footnote{By simple GUTs we mean anomaly free, 4D quantum field theory with a simple gauge group in the UV that breaks to the SM in the IR. We will also assume that there are no chiral fermions beyond the three SM generations.} in the literature contain or branch through $SU(5)$~\cite{Georgi:1974my,Fritzsch:1974nn,Gursey:1975ki,Georgi:1979md,Slansky:1981yr}, e.g. 
\begin{align}
    E_6 \supset SO(10) \supset SU(5) \, .
\end{align}
For all these cases, the SM gauge group is $[SU(3)\times SU(2)\times U(1)_Y] / \mathbb{Z}_6$~\cite{Tong:2017oea}. Both $SU(3)_c$ and $SU(2)_L$ are embedded at equal levels, the hypercharge normalization factor relative to QCD is fixed to be the standard 5/3.  The QCD axion couples to the photon with a coupling $E/N = 8/3$~\cite{Srednicki:1985xd,axion}. 

We present a new GUT which departs from this paradigm. The unified gauge group is $SU(12)$ (or $SU(12)/\mathbb{Z}_2$ depending on the matter content) and has QCD embedded at level $k_3= 3$, with $ k_2=1, k_Y = 1$. This particle content appears in the literature~\cite{atlas, anber}, but to our knowledge our work is the first GUT construction with a non-canonical SM embedding specified, c.f.~\cite{color,Slansky:1981yr,fonseca}. Our embedding has immediate consequences: the weak mixing angle at the GUT scale is $\sin^2 \theta_W = 1/2$ and the QCD axion-photon coupling is set by $E/N = 2/3$.

The $SU(12)$ example is the first member of an infinite family of embeddings into $SU(8m+4)$, with $m\ge1$. They share the ratio $E/N=2/3$ but have $\sin^2\theta_W(M_U)=3k_2/(4m+2)$, with $k_2=1$ or $3$ (the latter for $m\ge3$). Explicit anomaly-free fermion spectra with three SM generations are given in Sec.~\ref{app:su8m4-family} of the Supplemental Material. Remarkably, for $m=2$ we find that the theories do not need a 3-fold replication of UV fermions, but the minimal spectrum reproduces 3 SM generations, providing interesting candidates for flavor-unified GUTs. 

Interestingly, we find that for $SU(n)$ with $n\leq 37$ and any other compact simple Lie groups, $E/N = \{2/3, 8/3\}$ are the only two possibilities, irrespective of the dimension of the representations! This result is achieved by considering general constraints on the SM embedding in the GUT, a kind of GUT bootstrap that can scan over large classes of GUTs and SM embeddings. We provide further details in the appendix~\ref{app:gut-bootstrap} and a catalog of results as an application.

We comment on the novelty of our construction relative to the literature. Ref.~\cite{fonseca} scanned all SM embeddings in $SU(n)$ for $n\le12$ and, for $13\le n\le20$, those that go through an $SU(5)$ subgroup. Besides the standard embedding, the latter include $\mathbf{n}\to2\cdot\mathbf5+\bar{\mathbf5}+\dots$ for $n\ge15$, with $(k_3,k_2,k_Y)=(3,3,5)$, which corresponds to our family~II (Sec.~\ref{app:two-index-proof} of the Supplemental Material). All embeddings found there give $E/N=8/3$. Our $SU(12)$ embedding does not go through $SU(5)$ and is not found in the $n\le12$ scan of Ref.~\cite{fonseca}.
QCD at level 3 also appears in the color-flavor unified models of Refs.~\cite{Cordova:2024ypu,Delgado:2026lwj}, where $\mathbf9\to3\cdot\mathbf3$ under $SU(3)_c\subset SU(9)$, and in the $SU(12)_C\times SU(2)_L\times SU(2)_R$ models of Ref.~\cite{Li:2019sxg} (see also the classification in~\cite{Allanach:2021bfe}). In these models the weak group is not unified, so neither $\sin^2\theta_W$ nor $E/N$ is predicted.
Family unification using spinors of the groups $SO(4m+2)$~\cite{Wilczek:1981iz,Bagger:1984rk}, with $m>2$, provides an even number of SM generations. The additional standard and \textit{mirror} families must be removed from the spectrum~\cite{Giudice:1991sz,Reig:2017nrz}. 

In this paper we present the simplest version of the model, emphasizing the chiral content. 
There are a few interesting phenomenological consequences of this theory. If the gauge group is $SU(12)/\mathbb{Z}_2$, the SM gauge group is 
$SU(3)\times SU(2)\times U(1)/\mathbb{Z}_6$ and the usual electric
charge quantization follows~\cite{Tong:2017oea}. 
But if the group is $SU(12)$, we get a different global structure for the SM gauge group, $SU(3)\times SU(2)\times U(1)/\mathbb{Z}_3$, and vector-like fermions with $1/2$ charges are allowed.\footnote{See Ref.~\cite{Koren:2024xof} for a recent study of fractionally charged particles in the context of colliders.} 
For both $SU(12)$ and $SU(12)/\mathbb{Z}_2$ the smooth monopole carries two Dirac charge units as a consequence of the embedding, and there are no stable defects of higher dimension (strings or domain walls) associated to GUT symmetry breaking. 
$B+L$ is violated via dimension-6 operators similar to those in flipped-$SU(5)$ models and leads to proton decay that can be within current bounds~\cite{Super-Kamiokande:2020wjk} and could be distinguished from other GUTs. In contrast to GUTs based on $SO(10)$, there is no $B-L$ symmetry. 

The axion-photon coupling is one of the most distinctive phenomenological consequences and is what ultimately can distinguish this theory from standard GUTs. The theory contains fractional instantons that can have a non-trivial impact in axion physics, too.

There are many future model building directions -- characterizing the spectrum for different options for gauge coupling unification, the Higgs sector of the GUT (including, e.g. doublet-triplet splitting), flavor phenomenology (specially in the flavor unified models of App.~\ref{app:su8m4-family}), neutrino mass generation, and embedding in string theory. We note that the model can be made supersymmetric, but in this case the coupling runs strongly and leads to a Landau pole close to the unification scale.

\paragraph{\textbf{The model}. }The GUT gauge algebra is $\frak{g} = \frak{su}(12)$, and the embedding of the SM $\frak{h} = \frak{su}(3)\times \frak{su}(2) \times \frak{u}(1)_Y$ can be shown by the branching of the fundamental representation of $\frak{su}(12)$: 
\begin{align} \label{eq:fund_decomposition}
    \mathbf{12} \rightarrow
 2(3, 1)_{1/6}
 +(\bar{3},1)_{-1/6}
 +(1,2)_{0}
 +(1,1)_{-1/2}\,.
\end{align}
The indices of embedding are $(k_3,k_2,k_Y)=(3, 1, 1)$  and can be obtained from the normalization of the generators~\cite{Dynkin:1957um,Slansky:1981yr}
\begin{align}
\left
(\operatorname{tr}_VT_c^2,\operatorname{tr}_V(T_L^3)^2,
\operatorname{tr}_VY^2\right)=(3/2,1/2,1/2).
\end{align}
We will use the indices $i = 1, \ldots, 5$ to label the 5 SM representations in the branching of the $\rep{12}$ in eq.~\eqref{eq:fund_decomposition}.

The fermion content organizes into three copies of the antisymmetric and (conjugate) symmetric two-index representations, $2\times\mathbf{66}+\overline{\mathbf{78}}$. Each of these contains exactly one SM family together with vector-like representations (denoted $\mathcal{S}$, see App.~\ref{app:branching} for details),
\begin{equation}\label{eq:family_main}   2\times\mathbf{66}+\overline{\mathbf{78}}\;\to\;\underbrace{Q+u^c+d^c+L+e^c}_{\text{one SM family}}\;+\;\mathcal{S}\,.
\end{equation}
The $\mathcal{S}$ contains fermions that all admit SM-invariant masses, so the only chiral states below $M_U$ are the three SM families.

The $\mathbf{66}$ and $\overline{\mathbf{78}}$ representations have even 12-ality, which allows this model to have gauge group $SU(12)$ or $SU(12)/\mathbb{Z}_2$. Representations with even 12-ality have only SM-like charges, with integer electric charges. However, if the gauge group is $SU(12)$ it allows for fermions in the fundamental $\mathbf{12}$ representation.  In that case, the low energy gauge group is
\begin{equation}
    [SU(3)\times SU(2)\times U(1)]/\mathbb{Z}_3\,,
\end{equation}
and every color-singlet state from the fundamental has an electric charge which is an integer multiple of $1/2$. This is therefore a GUT that would survive the discovery of exotic particles with fractional charges. 

\paragraph{\textbf{Predictions.}} The standard GUT predictions are modified due to the non-standard embedding of QCD. Below we briefly discuss the most salient features.

\begin{figure}[t]
\centering
\includegraphics[width=\columnwidth]{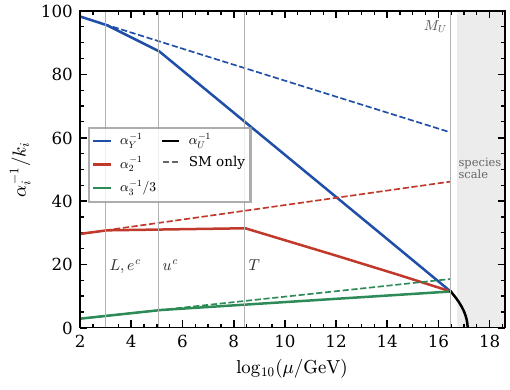}
\caption{Two-loop running of $\alpha_i^{-1}/k_i$, with $k=(3,1,1)$ for $M_U=3\times10^{16}$~GeV with one added $\mathbf{78}+\overline{\mathbf{78}}$ pair. Solid: with the vector-like states $L,e^c$, $u^c$ and $T$ (gray lines mark their masses); dashed: SM content only. Inputs at $M_Z$ from Ref.~\cite{ParticleDataGroup:2024cfk}. Above $M_U$ (black) the $SU(12)$ coupling grows and reaches a formal pole at $\simeq4.6M_U$; the gray band lies above the species scale $\Lambda_{\rm sp}$~\cite{Dvali:2007hz}.}
\label{fig:running}
\end{figure}

\paragraph{The weak mixing angle.}---The gauge coupling matching condition is
\begin{equation}\label{eq:matching_condition}
    \alpha^{-1}_i(M_U)=k_i \alpha^{-1}_{\rm GUT}\,.
\end{equation}
The weak mixing angle at the GUT scale can also be written in terms of the levels of embedding as~\cite{Ginsparg:1987ee,axion}
\begin{equation}\label{eq:sin_theta}
    \sin^2\theta_W(M_U)=\frac{k_2}{k_Y+k_2}=\frac{1}{2}\,.
\end{equation}
This is to be compared with $3/8$ for standard GUT embeddings. 
This requires states below the unification scale to modify the running from the SM (or MSSM) predictions, see Fig.~\ref{fig:running}. We give some representative examples of how this can be achieved below.

\paragraph{The axion-photon coupling.}---The ratio of the anomaly coefficients, which sets the axion-photon coupling, can be written in terms of the levels of embedding~\cite{axion}
\begin{equation}\label{eq:E/N}
\frac EN=\frac{k_Y+k_2}{k_3}=\frac{2}{3}\,.
\end{equation}

\begin{figure}[t]
\centering
\includegraphics[width=\columnwidth]{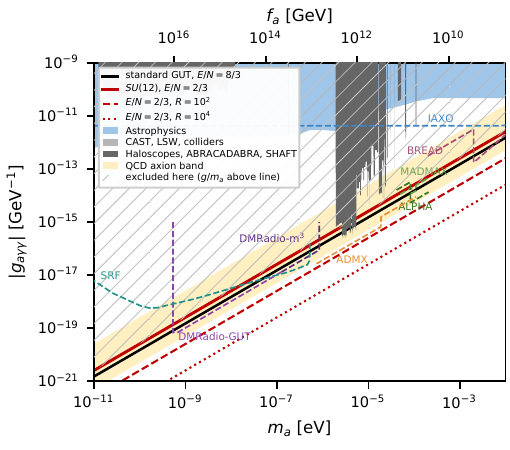}
\caption{Axion-photon coupling versus mass. Black: standard GUTs, $E/N=8/3$. Solid red: this model, $E/N=2/3$, with a QCD-dominated axion mass. Dashed and dotted red: the same with an aligned contribution from small $SU(12)$ instantons, $m_a^2f_a^2=\chi(1+R)$, for $R=10^2$ and $10^4$. Hatched: not allowed in this model. As in standard GUTs axion-like particles above the QCD line are forbidden~\cite{axion}. Yellow: the QCD axion band, $0.25<|E/N-1.92|<12.75$~\cite{DiLuzio:2016sbl}. Shaded regions and dashed curves: limits and projections from Ref.~\cite{AxionLimits}.}
\label{fig:axion}
\end{figure}

The axion coupling to photons is given by~\cite{preciseaxion}
\begin{equation}\label{eq:gagg}
   g_{a\gamma\gamma}=\frac{\alpha_{\rm em}}{2\pi f_a}\left(\frac EN-1.92\right),
\end{equation}
so that for a fixed decay constant, $E/N=2/3$ gives a coupling 1.7 times larger than the standard GUT value $E/N=8/3$ (see Fig.~\ref{fig:axion}).

This provides a well-motivated target for experimental precision. The power in a haloscope experiment scales as $g_{a\gamma\gamma}^2\rho_a$. If a DM axion is detected, and its mass is measured accurately, the two embeddings differ by a factor $\sim 2.9$ in the power. Distinguishing the non-standard GUT studied here from canonical unified models requires reaching sufficient precision, usually limited by the determination of the local DM abundance. The Gaia experiment may reduce it to $10$\% in a near future~\cite{Lim:2023lss,Putney:2025clr}. 

\paragraph{Gauge-coupling unification.}---The prediction~\eqref{eq:sin_theta} is far from the SM or MSSM extrapolation. The running is sensitive to the details of the spectrum. Matching the gauge couplings requires $\alpha_Y^{-1}$ to decrease faster than the SM predicts, requiring light states with hypercharge, as well as new colored states that slow down the evolution of the QCD coupling. 

We find two representative options. With the minimal content, $6\times\mathbf{66}+3\times\overline{\mathbf{78}}$, a one-loop bound allowing arbitrary partial multiplets gives $\alpha_U^{-1}>0$ only for $M_U>1.75\times10^{17}$~GeV, and $\alpha_U^{-1}\simeq8$--$12$ requires $M_U\simeq10^{18}$--$2.4\times10^{18}$~GeV, that is a GUT scale close to the Planck scale. Adding one vector-like pair $\mathbf{78}+\overline{\mathbf{78}}$ allows unification at $M_U=3\times10^{16}$~GeV with $\alpha_U^{-1}\simeq11.4$ at two loops. For this benchmark there exist new degrees of freedom at low energies, 4 Dirac pairs of  $L=(\mathbf{1},\mathbf{2})_{-1/2}$ and one Dirac $e^c=(\mathbf{1},\mathbf{1})_{1}$ at $1$~TeV, 4 Dirac $u^c=(\mathbf{\overline{3}},\mathbf{1})_{-2/3}$ at $\sim10^{5}$~GeV and 5 Weyl $T=(\mathbf{1},\mathbf{3})_0$ at $\sim3\times10^{8}$~GeV (see Fig.~\ref{fig:running}). In both cases the remaining states are at $M_U$. Above $M_U$ the large matter content makes the $SU(12)$ gauge coupling grow fast, and the theory must be completed within a factor of a few of $M_U$. This is comparable to the species scale $\Lambda_{\rm sp}\sim \bar M_{\rm Pl}/\sqrt{N_{\rm sp}}\sim 2M_U$~\cite{Dvali:2007hz}. 

Unlike standard GUTs where the weak mixing angle is a successful prediction of unification, here it is an aspect that requires light vectorlike exotic states, some of which lie around the TeV scale and provide well-motivated collider benchmarks.

\paragraph{The scalar sector.}---
A minimal scalar sector that suffices to break the GUT down to the SM contains a real adjoint $P\in\rep{143}$, a complex
$K\in\rep{66}$, and a complex singlet $\phi$ (responsible for PQ breaking in models with an axion). 
$SU(12)$ is broken to the SM by the vacuum expectation value (vev)
\begin{equation}\label{eq:vevs}
\begin{aligned}
    \vev{P}&=v\,\text{diag}\big(-3\mathbf{1}_3,\,2\mathbf{1}_3,\,-\mathbf{1}_3,\,3\mathbf{1}_2,\,0\big),\\
    \vev{K_{13}}&=\vev{K_{23}}=v\,\mathbf{1}_3,\qquad \vev{K_{44}}=v\,\epsilon_2 ,
\end{aligned}
\end{equation}
and $\vev{\phi}=v$.
The indices $i, j = 1, \ldots, 5$ label the SM representations that make up the $\rep{12}$ of $SU(12)$ in eq.~\eqref{eq:fund_decomposition}. A more detailed analysis of the possible scalar sectors and the relations between fermion masses that result is beyond the scope of this letter.

$P$ and $K$ each contain one $(\rep1,\rep2)_{1/2}$ doublet which we call $H_P\sim P^4{}_5$ and $H_K\sim(K^{45})^*$. One combination
is eaten by the $(\rep1,\rep2)_{1/2}$ gauge boson, and the orthogonal one is the  candidate for the SM Higgs,
\begin{equation}
H_{\rm SM}=c_PH_P+c_KH_K.
\label{eq:higgs}
\end{equation}
The eaten combination is fixed by the broken generator in the $(\rep1,\rep2)_{1/2}$ direction acting on the vevs. We find $c_P^2=2/11$ for the vevs in Eq.~\eqref{eq:vevs}. More generally, for $\vev{P}\propto v_P$ and $\vev{K}\propto v_K$, $c_P^2=2v_K^2/(9v_P^2+2v_K^2)$. Only $H_P$ has renormalizable Yukawa couplings.  Giving all vectorlike spectators masses requires additional scalar representations or higher-dimensional operators; the physical SM Yukawas then depend on the resulting Higgs couplings and light-fermion mixing.
The mass of $H_{\rm SM}$ is not protected and must be tuned, as in standard GUTs. It would be nice to explore if this class of theories allows for new solutions to the analog of the doublet-triplet splitting problems of standard GUTs, and explore cases where the doublet arises as a pseudo-Goldstone.

\paragraph{Topological defects.}---
An element of $SU(12)$ that commutes with $\vev P$ is block diagonal, $g=\mathrm{diag}(g_1,\dots,g_5)$. Invariance of
$\vev{K_{13}}$ and $\vev{K_{23}}$ requires $g_1=g_2\equiv u$ and $g_3=u^*$, invariance of $\vev{K_{44}}$ requires
$\det g_4=1$, and $\det g=1$ fixes $g_5$. The unbroken subgroup of $SU(12)$ is therefore exactly
\begin{equation}
H=\big\{\mathrm{diag}\big(u,\,u,\,u^*,\,w,\,(\det u)^{-1}\big)\big\}\cong U(3)\times SU(2),
\label{eq:H}
\end{equation}
for $u \in U(3), w\in SU(2)$, hypercharge the abelian factor of $U(3)$, and $SU(2)_L$ a direct factor. $H$ is connected and
$\pi_1(H)=\mathbb{Z}$. The breaking therefore produces monopoles, $\pi_2(G/H)=\mathbb{Z}$, but no topologically
stable strings or domain walls, $\pi_1(G/H)=\pi_0(G/H)=0$. The same holds for $G=SU(12)/\mathbb{Z}_2$, where
$H=[U(3)\times SU(2)]/\mathbb{Z}_2$ is the SM group with its $\mathbb{Z}_6$ quotient. In this respect the model
resembles $SU(5)$ rather than the $SO(10)$, as in the latter monopoles can be attached by $\mathbb{Z}_2$ strings and domain walls can be bounded by
strings~\cite{Dunsky:2021tih,Buchmuller:2023aus}. 

The generator of $\pi_1(H)$ is the loop $u=\mathrm{diag}(1,1,e^{it})$, $0\le t\le2\pi$, with magnetic generator
\begin{equation}
Q_M=\mathrm{diag}(0,0,1\,|\,0,0,1\,|\,0,0,-1\,|\,0,0\,|\,-1)=2Y+\lambda_c ,
\label{eq:QM}
\end{equation}
where $\lambda_c=\mathrm{diag}(-\tfrac13,-\tfrac13,\tfrac23)$ on a color triplet. It winds one third of the
hypercharge circle $e^{6iY\theta}$, twice the minimal loop of the SM group (with $\Gamma = \mathbb{Z}_6$). The lightest 't~Hooft--Polyakov monopole
thus carries two Dirac units, $g=4\pi/e$, together with $\mathbb{Z}_3$ color-magnetic charge and no weak charge:
the long-range field of the doubly charged monopole of $SU(5)$. There is no singly charged soliton, so it is stable within the field theory. 

The global form determines the remaining one-form symmetry~\cite{Gaiotto:2014kfa,Aharony:2013hda,Tong:2017oea}. For $G=SU(12)$ the minimal charge is $e/2$, and the doubly charged monopole saturates Dirac quantization. For $SU(12)/\mathbb{Z}_2$, Dirac quantization allows $g=2\pi/e$, but there is no smooth field configuration with that charge. Since smooth monopoles have $g=4\pi/e$ there is a magnetic $\mathbb{Z}_2^{(1)}$ one-form symmetry in this case. Completeness of the spectrum~\cite{Polchinski:2003bq} suggests that this symmetry is broken in the full gravitational theory and singly charged monopoles exist in the UV.

\paragraph{Proton decay channels.}---
Gauge bosons live in the adjoint representation. From the branching of the fundamental we see that the adjoint of $SU(12)$ will contain a $(\mathbf{3},\mathbf{2})_{1/6}$ but no $(\mathbf{3},\mathbf{2})_{-5/6}$. This differs from standard $SU(5)$ and $SO(10)$ GUTs~\cite{Georgi:1974sy,Georgi:1974my,Fritzsch:1974nn} and resembles flipped $SU(5)$~\cite{Barr:1981qv,Derendinger:1983aj,Antoniadis:1987dx}. Tree-level gauge boson exchange then generates the same operators as in the latter~\cite{Dorsner:2004xx,Ellis:2020qad}, that is $QL\,u^{c\dagger}d^{c\dagger}$ but no operator with $e^c$. The dominant proton decay modes are $p\to\bar\nu\pi^+$ and $p\to\ell^+\pi^0$, with $\Gamma(\bar\nu\pi^+)\gtrsim2\,\Gamma(e^+\pi^0)$, depending on lepton mixing, see~\cite{Dorsner:2004xx,Ellis:2020qad}. In addition to the gauge-mediated processes there will be (model-dependent) scalar-mediated processes from the representation that breaks $SU(12)$ down to the SM. 

\paragraph{Fractional instantons and the QCD axion.}---Because the unified coupling is stronger than usual, GUT instantons are less suppressed than in normal GUTs~\cite{Holdom:1982ew,Agrawal:2017ksf}. Their tree level action is
\begin{align}
    S=\frac{2\pi}{\alpha_U}=\frac{2\pi}{k_i\alpha_i}\,,
\end{align}
and might spoil the axion solution to the strong CP problem~\cite{Kamionkowski:1992mf,Holman:1992us,Barr:1992qq}. 

For the $SU(12)$ theory, the instanton has an action $S=2\pi/(3\alpha_{\rm s})$. Because QCD sits in $SU(12)$ at level 3, a unit $SU(12)$ instanton carries one third of the QCD topological charge. Fractional instantons near the GUT scale contribute to the axion potential as
\begin{align}
    &V(a) = -\chi\cos\theta - \Lambda_U^4\cos (\theta/3 +\delta),\,\,\,\,\,  \\&m_a^2f^2_a = \chi(1+R),\,\,\,\,\,R=\Lambda_U^4\cos\delta/(9\chi)\,.
\end{align}
Here $\delta $ is a model-dependent CP-violating phase indicating that the two contributions are not necessarily aligned. For $\delta\sim 1$, $|\bar\theta|\lesssim 10^{-10}$ requires $\Lambda_U^4/\chi\lesssim 10^{-10}$. The scale $\Lambda_U$ follows from the constrained instanton density~\cite{Affleck:1980mp,Fuentes-Martin:2019bue,Csaki:2019vte},
\begin{equation}
    \Lambda_U^4\simeq \,C_{12}\!\int\!\frac{d\rho}{\rho^5}\,K_U\Big(\frac{2\pi}{\alpha_U(1/\rho)}\Big)^{24}e^{-2\pi/\alpha_U(1/\rho)}\,e^{-\pi^2\rho^2v^2}\,,
\end{equation}
where $C_{12}$ is a calculable constant. If the $SU(12)$ theory is not asymptotically free above $M_U$, the integral may be dominated by the smallest instantons, $\rho^{-1}\sim$ a few $M_U$. The factor $K_U$ accounts for the fermion zero modes of the 't Hooft vertex, which must be lifted by fermion masses or Yukawa couplings. Since the $SU(12)$ content above carries many zero modes, $K_U$, and hence $\Lambda_U$, is very sensitive to the heavy vector-like spectrum. For $R\gg 1$, the axion is heavier by $\sim \sqrt{R}$ and the QCD line moves towards smaller $g_{a\gamma\gamma}/m_a$.

\paragraph{\textbf{Towards flavor unification.}}--- In the $SU(12)$ model the three families come from three copies of $2\,\mathbf{66}+\overline{\mathbf{78}}$, as in standard GUTs. This is no longer required for larger members of the $SU(8m+4)$ family (see.~\ref{app:su8m4-family}). Anomaly cancellation fixes the fermion content to $(m+1)$ antisymmetric and $m$ conjugate symmetric representations, and for some embeddings this minimal spectrum already contains exactly three SM families (see.~\ref{app:su8m4-family}). The first example is $SU(20)$ with $3\,\times \,\mathbf{190}+2\,\times \,\overline{\mathbf{210}}$. Its embedding indices are $(k_3,k_2,k_Y)=(5,1,7/3)$, which give $E/N=2/3$ and $\sin^2\theta_W(M_U)=3/10$. In these theories the number of generations is fixed by the embedding rather than by replication, making them promising candidates for gauge and flavor unification.

\paragraph{\textbf{Conclusions.}}---We presented the first 4D GUT in which the SM group does not factor through $SU(5)$. The embedding is characterized by the integers $(k_3,k_2,k_Y)=(3,1,1)$ which control the main predictions. The SM spectrum arises from $6\,\mathbf{66}+3\,\overline{\mathbf{78}}$ of $SU(12)$ and contains only three chiral families and vector-like matter.

The theory is characterized by the prediction $\sin^2\theta_w = 1/2$ and the absence of $(\mathbf{3},\mathbf{2})_{-5/6}$ gauge bosons. Proton decay occurs in a manner similar to flipped $SU(5)$ models. Minimal models achieve gauge coupling unification provided that there are vector-like states near the TeV scale. In these models, there are fermions transforming as $L=(\mathbf{1},\mathbf{2})_{-1/2}$ within the reach of HL-LHC and other future colliders, as shown in Fig.~\ref{fig:running}. In the case the gauge group is $SU(12)$ instead of $SU(12)/\mathbb{Z}_2$, the fundamental representation is allowed, and color-singlet states carry half of the electron charge. A particle of charge $\pm e/2$ would then be consistent with this class of models, but charge $\pm e/3$ would rule it out. The monopole spectrum contains only doubly charged smooth monopoles.

A robust prediction is the axion-photon coupling, determined by the ratio $E/N=2/3$, rather than the prediction $E/N=8/3$ which holds for GUTs which branch through $SU(5)$. We have shown that $E/N=2/3$ is the only possible alternative to the standard $E/N=8/3$ for GUTs with no chiral exotica, up to $SU(n)$ models with $n>37$. In these theories, $E/N$ is a sharper test of unification than proton decay, as a value outside of $E/N = \{2/3, 8/3\}$ would rule out GUT models without chiral exotica and $n<37$. The axion-photon coupling is a factor 1.7 larger if $E/N = 2/3$ relative to $E/N=8/3$, which is potentially measurable in haloscope experiments in the event of an axion discovery and an improved estimate of the local dark matter density. As in standard GUTs an axion above the QCD line, would exclude it, too~\cite{axion}.

The non-standard embedding also has implications for the strong CP problem. Because $\alpha_U=3\alpha_s(M_U)$, small $SU(12)$ instantons are less suppressed than standard GUTs. Their contribution depends strongly on the heavy spectrum, and how fermionic zero modes are saturated. These fractional instantons can spoil the solution of the strong CP problem unless they are subdominant or aligned with QCD. They can also easily make the QCD axion much heavier than standard predictions.

The $SU(12)$ theory that we studied in detail is the first member of an infinite family based on $SU(8m+4)$. We have shown that using up to two-index representations, we can reduce all $SU(n)$ embeddings to the canonical $SU(5)$ embedding or our new embedding. We have also developed a simple GUT boostrap, which explores the space possible GUTs and SM embeddings using general constraints, independent of representation.  It is exciting to see that some of these, as the $SU(20)$ theory we briefly discuss in \ref{app:su8m4-family}, generate three SM families from minimal UV spectra without flavor replication.

Several interesting directions remain. A detailed study of the symmetry breaking pattern classifying intermediate phases, a complete calculation of the impact of small instantons on the QCD axion, as well as neutrino mass generation by the singlets or flavor predictions (e.g. $b-\tau$ unificaxion) are interesting follow-up directions. 

\begin{acknowledgments}
The catalog in the appendix was built with extensive help of ChatGPT Astra, including generating the set of constraints for the bootstrap and the proof for the two-index theorem. The driving question was a group theory classification for the possible QCD axion-photon couplings. The novel embedding was found by the LLM as a counterexample to an $E/N=8/3$ theorem. All explicit results and constraints have been checked by the authors.
ChatGPT and Claude were used to help edit and check the manuscript; however it is almost entirely written by --- the authors.

The research of PA is supported by the U.S. Department of Energy
under the grant DE-SC0011702.
\end{acknowledgments}

\clearpage
\onecolumngrid
\begin{center}
{\large\bfseries Supplemental Material}
\end{center}

\setcounter{equation}{0}
\setcounter{figure}{0}
\setcounter{table}{0}
\setcounter{section}{0}
\setcounter{secnumdepth}{1}
\renewcommand{\theequation}{S\arabic{equation}}
\renewcommand{\thefigure}{S\arabic{figure}}
\renewcommand{\thetable}{S\arabic{table}}
\renewcommand{\thesection}{S\arabic{section}}
\renewcommand{\theHequation}{supp.\arabic{equation}}
\renewcommand{\theHfigure}{supp.\arabic{figure}}
\renewcommand{\theHtable}{supp.\arabic{table}}
\renewcommand{\theHsection}{supp.\arabic{section}}

\section{Branching and particle content for $SU(12)$}\label{app:branching}
In this appendix we provide the detailed spectrum for the $SU(12)$ branching in the main text. 
We label the familiar SM representations of left-handed Weyl fermions as $Q, u^c, d^c, L, e^c$.
In table~\ref{tab:sm-multiplets} we show the quantum numbers of the SM fermions, the spectator multiplets, and other multiplets that appear in the defining blocks used below.
Bars denote conjugate representations (e.g. $\overline Q = (\bar{3}, 2)_{-1/6}$).

\begin{table}[h!]
\centering
\begingroup
\small
\renewcommand{\arraystretch}{1.2}
\begin{tabular}{@{}cl@{\hspace{2em}}cl@{\hspace{2em}}cl@{}}
\toprule
\multicolumn{2}{c}{SM fermions} & \multicolumn{2}{c}{Spectator multiplets}
 & \multicolumn{2}{c}{Defining blocks}\\
\cmidrule(lr){1-2}\cmidrule(lr){3-4}\cmidrule(lr){5-6}
Label & Representation & Label & Representation & Label & Representation\\
\midrule
$Q$ & $(\rep3,\rep2)_{1/6}$
 & $X$ & $(\rep6,\rep1)_{1/3}$
 & $C$ & $(\rep3,\rep1)_{1/6}$\\
$u^c$ & $(\overline{\rep3},\rep1)_{-2/3}$
 & $O$ & $(\rep8,\rep1)_0$
 & $D$ & $(\rep1,\rep2)_0$\\
$d^c$ & $(\overline{\rep3},\rep1)_{1/3}$
 & $T$ & $(\rep1,\rep3)_0$
 & $\sigma$ & $(\rep1,\rep1)_{-1/2}$\\
$L$ & $(\rep1,\rep2)_{-1/2}$
 & $N$ & $(\rep1,\rep1)_0$
 & $F_5$ & $(\rep3,\rep1)_{-1/3}+(\rep1,\rep2)_{1/2}$\\
$e^c$ & $(\rep1,\rep1)_1$ & & & &\\
\bottomrule
\end{tabular}
\endgroup
\caption{Notation for representations of $SU(3)_c\times SU(2)_L\times U(1)_Y$.
The spectator representations $O,T,N$ are real.}
\label{tab:sm-multiplets}
\end{table}

The defining embedding in Eq.~\eqref{eq:fund_decomposition} and its
antisymmetric and conjugate symmetric two-index branchings are shown
in Table~\ref{tab:index-blocks}. Here $t_3$ and $t_2$ are arbitrary
traceless Hermitian matrices in the fundamental representations of
$SU(3)_c$ and $SU(2)_L$, respectively; $I_3$ is the $3\times3$ identity,
and $y$ a real coefficient multiplying
hypercharge.

\begin{table}[h!]
\centering
{\small
\setlength{\arraycolsep}{6pt}
\renewcommand{\arraystretch}{1.15}
\[
X_{\rm SM}=
\begin{pmatrix}
t_3+\frac{y}{6}I_3 & 0 & 0 & 0 & 0 \\
0 & t_3+\frac{y}{6}I_3 & 0 & 0 & 0 \\
0 & 0 & -t_3^{\mathsf T}-\frac{y}{6}I_3 & 0 & 0 \\
0 & 0 & 0 & t_2 & 0 \\
0 & 0 & 0 & 0 & -\frac{y}{2}
\end{pmatrix}.
\]
}
\par\medskip
\begin{minipage}[t]{0.49\textwidth}
\centering
\small
\setlength{\arraycolsep}{6pt}
\renewcommand{\arraystretch}{1.15}
$A\in\mathbf{66}$
\[
\begin{pmatrix}
\mathsf A_3[d^c] & \mathsf A_3[d^c]+\mathsf S_3[X] & O+N & Q & \overline{d^c} \\
* & \mathsf A_3[d^c] & O+N & Q & \overline{d^c} \\
* & * & \mathsf A_3[\overline{d^c}] & \overline Q & u^c \\
* & * & * & \mathsf A_2[N] & L \\
* & * & * & * & 0
\end{pmatrix}
\]
\end{minipage}\hfill
\begin{minipage}[t]{0.49\textwidth}
\centering
\small
\setlength{\arraycolsep}{6pt}
\renewcommand{\arraystretch}{1.15}
$\overline{S}\in\overline{\mathbf{78}}$
\[
\begin{pmatrix}
\mathsf S_3[\overline X] & \mathsf A_3[\overline{d^c}]+\mathsf S_3[\overline X] & O+N & \overline Q & d^c \\
* & \mathsf S_3[\overline X] & O+N & \overline Q & d^c \\
* & * & \mathsf S_3[X] & Q & \overline{u^c} \\
* & * & * & \mathsf S_2[T] & \overline L \\
* & * & * & * & e^c
\end{pmatrix}
\]
\end{minipage}
\caption{Defining embedding (top) and schematic tensor block decompositions.
$\mathsf A_r[F]$ and $\mathsf S_r[F]$ are the antisymmetric and symmetric
$r\times r$ matrices built linearly from the components of $F$ in the
relevant block bases. Thus the triplet--triplet block contains
$\mathsf A_3[d^c]+\mathsf S_3[X]$.
Other entries label the SM representation content; repeated labels in
different blocks denote independent field copies. Stars are fixed by
$A_{JI}=-(A_{IJ})^{\mathsf T}$ and $\overline S_{JI}=(\overline S_{IJ})^{\mathsf T}$,
where $I,J$ label the five defining blocks.}
\label{tab:index-blocks}
\end{table}

For the spectrum $\mathcal{F} = 2\times\rep{66}+\overline{\rep{78}}$,
the net spectrum per UV family branches to
\begin{align}
2\times\rep{66}+\overline{\rep{78}}\downarrow H
&=\underbrace{Q+u^c+d^c+L+e^c}_{\text{one SM family}}+\mathcal S,
\label{eq:family}\\
\mathcal S&=4(Q+\overline Q)+(u^c+\overline u^c)+7(d^c+\overline d^c)
\nonumber\\
&\qquad
+(L+\overline L)+3(X+\overline X)+6O+T+8N.
\nonumber
\end{align}
This corresponds to  $210=15+195$ Weyl components per family, or $630=45+585$ in total.
All spectators in $\mathcal{S}$, including $N$, admit SM-preserving masses.

\section{Five Families of $SU(n)$ embeddings}\label{app:su8m4-family}
The $SU(12)$ example in the text is a special case of a family of SM embeddings in $SU(n)$ for $n=8m+4$, $m\geq 1$. We will show a proof in section~\ref{app:two-index-proof} that there are only five families of $SU(n)$ embedding up to two-index representations. Two of these are the canonical $SU(5)$ embeddings, and the other three families are based on $SU(8m+4)$ and are closely related to the $SU(12)$ embedding in the text.
The $SU(n)$ representations of interest are
$A_2=\wedge^2\rep n$ and $S_2=\operatorname{Sym}^2\rep n$.
All fermions below are left-handed Weyl fields. The construction continues
to arbitrarily large $n$.

\paragraph{Embedding and predictions.}
The $SU(8m+4)$ family of embedding is defined by 3 classes labeled by $(k,d)\in\{(1,1),(1,3),(3,1)\}$.  Setting
$a=m+(d+1)/2$, $b=m+(1-d)/2$, with $a,b\ge k$, 
the defining representation branches as
\begin{equation}
 \begin{aligned}
 V=\rep n\ \longrightarrow\quad &
 aC+b\overline C+kD+(a-k)\sigma+(b-k)\overline\sigma.
 \end{aligned}
 \label{eq:series-defining}
\end{equation}
Thus $k$ counts weak doublets and $d=a-b$ is the triplet--antitriplet
imbalance in $V$. The sum of dimensions of the SM representations dimension is $n$ and $\Tr_VY=0$.
The physical hypercharge is defined as $Q=T_L^3+Y$ and nonabelian fundamental
normalization $\Tr T^a T^b = \delta^{ab} / 2$. 
The embedding indices are
\begin{equation}
 (k_3,k_2,k_Y)=\left(2m+1,\ k,\ \frac{4m+2}{3}-k\right).
 \label{eq:series-indices}
\end{equation}
For a PQ symmetry commuting with the unified group and having a nonzero
mixed gauge anomaly, and with tree-level gauge matching at $M_U$,
\begin{equation}
 \boxed{\frac EN=\frac{k_Y+k_2}{k_3}=\frac23},\qquad
 \boxed{\sin^2\theta_W(M_U)=\frac{k_2}{k_Y+k_2}=\frac{3k}{4m+2}}.
 \label{eq:series-predictions}
\end{equation}

\paragraph{Three-generation fermion spectra.}
The fermion content for the GUT is
\begin{equation}
 \mathcal F=\frac{3}{dk}\left[(m+1)A_2+m\overline S_2\right].
 \label{eq:series-fermions}
\end{equation}
The three choices above give nonnegative integral multiplicities. Along with the $SU(5)$ embeddings, we get five families, shown in table~\ref{tab:two-index-families}.

\begin{table}[h!]
\centering
\begingroup
\small
\setlength{\tabcolsep}{5pt}
\renewcommand{\arraystretch}{1.25}
\begin{tabular}{@{}ccllcc@{}}
\toprule
Family & $n$ & $V\downarrow_{\rm SM}$ & $\mathcal F$
 & $(k_3,k_2,k_Y)$ & $E/N$\\
\midrule
I & $5$
 & $F_5$ & $3A_2+3\overline V$ & $(1,1,5/3)$ & $8/3$\\[6pt]
\midrule
II & $15$
 & $2F_5+\overline F_5$ & $2A_2+\overline S_2+3\overline V$
 & $(3,3,5)$ & $8/3$\\[6pt]
\midrule
III & $8m+4$
 & $(m+1)C+m\overline C+D+m\sigma+(m-1)\overline\sigma$
 & $3\bigl[(m+1)A_2+m\overline S_2\bigr]$
 & $\bigl(2m+1,1,\tfrac{4m-1}{3}\bigr)$ & $2/3$\\[6pt]
\midrule
IV & $8m+4$
 & $(m+2)C+(m-1)\overline C+D+(m+1)\sigma+(m-2)\overline\sigma$
 & $(m+1)A_2+m\overline S_2$
 & $\bigl(2m+1,1,\tfrac{4m-1}{3}\bigr)$ & $2/3$\\[6pt]
\midrule
V & $8m+4$
 & $(m+1)C+m\overline C+3D+(m-2)\sigma+(m-3)\overline\sigma$
 & $(m+1)A_2+m\overline S_2$
 & $\bigl(2m+1,3,\tfrac{4m-7}{3}\bigr)$ & $2/3$\\[6pt]
\bottomrule
\end{tabular}
\endgroup
\caption{Five families of two-index embeddings in $SU(n)$, with defining
branching $V$ and UV Weyl-fermion spectrum $\mathcal F$.
The defining blocks are listed in Table~\ref{tab:sm-multiplets}. Family II corresponds to the higher-level-embedding found in ref.~\cite{fonseca}.
Families III, IV and V have $m\ge1$, $m\ge2$ and $m\ge3$,
and correspond to $(k,d)=(1,1),(1,3),(3,1)$, respectively.
The table displays cores without neutral defining singlets.}
\label{tab:two-index-families}
\end{table}

The last two rows use different SM embeddings despite identical GUT
fermion representations.
It is worth emphasizing that families IV and V do not have a 3-fold replication, with the three SM generations coming from one copy of $\mathcal{F}$ in the UV theory. The smallest anomaly-free spectrum with $A_2, S_2$ with this embedding naturally produces 3 generations of the SM; hence, these constitute a model of flavor unification as well! The smallest group for which this works is $SU(20)$.

\paragraph{Anomaly cancellation:}
Normalizing the cubic anomaly of $\rep n$ to
one, the anomaly for these reps are $\mathcal A_G(A_2)=n-4$ and $\mathcal A_G(\overline S_2)=-(n+4)$, so $\mathcal{F}$ satisfies anomaly cancellation,
\begin{equation}
 \mathcal A_G(\mathcal F)=\frac{3}{dk}
 \big[(m+1)(n-4)-m(n+4)\big]=0.
 \label{eq:series-anomaly}
\end{equation}

\paragraph{Chirality} The branching of $A_2$ and $S_2$ to the SM can be calculated by using
\begin{equation}
\begin{aligned}
\wedge^2 \left(\bigoplus c_i R_i\right)
={}& \bigoplus_i c_i\,\wedge^2 R_i
\oplus \bigoplus_i \binom{c_i}{2}(R_i\otimes R_i)
\oplus \bigoplus_{i<j}c_ic_j(R_i\otimes R_j).
\end{aligned}
\end{equation}
The same formula holds for the symmetric square upon replacing
\(\wedge^2\) by \(\operatorname{Sym}^2\) on both sides.

The branchings can be written in terms of the SM representations defined in appendix~\ref{app:branching}. The expression is,
\begin{equation}
\begin{aligned}
A_2\downarrow_{\mathrm{SM}}={}&
akQ+bk\overline Q
+b(a-k)u^c+a(b-k)\overline{u^c}
+\left[\binom{a+1}{2}+b(b-k)\right]d^c
+\left[\binom{b+1}{2}+a(a-k)\right]\overline{d^c}\\
&+k(a-k)L+k(b-k)\overline L
+\binom{b-k}{2}e^c+\binom{a-k}{2}\overline{e^c}
+\binom a2 X+\binom b2\overline X\\
&+abO+\binom k2 T
+\left[ab+(a-k)(b-k)+\binom{k+1}{2}\right]N .
\end{aligned}
\label{eq:series-antisymmetric-branching}
\end{equation}

\begin{equation}
\begin{aligned}
\overline S_2\downarrow_{\mathrm{SM}}={}&
bkQ+ak\overline Q
+a(b-k)u^c+b(a-k)\overline{u^c}
+\left[\binom b2+a(a-k)\right]d^c
+\left[\binom a2+b(b-k)\right]\overline{d^c}\\
&+k(b-k)L+k(a-k)\overline L
+\binom{a-k+1}{2}e^c+\binom{b-k+1}{2}\overline{e^c}
+\binom{b+1}{2}X+\binom{a+1}{2}\overline X\\&+abO+\binom{k+1}{2}T
+\left[ab+(a-k)(b-k)+\binom k2\right]N .
\end{aligned}
\label{eq:series-symmetric-branching}
\end{equation}

It is straightforward to check that the net chiral spectrum $\Delta R \equiv R - \overline{R}$ gives the SM,
\begin{equation}
 \Delta\!\left[(m+1)A_2
 +m \, \overline S_2\right]
 =dk\, \Delta \mathcal F_{\rm SM},
 \label{eq:series-chirality}
\end{equation}
where $\mathcal F_{\rm SM}=Q+u^c+d^c+L+e^c$ is one SM generation.
All other complex representations pair with their conjugates. Therefore those, as well as the real spectators $O$,
$T$ and $N$ all admit a SM-preserving mass. Since $dk = 3$ for three SM families, we see that $SU(2)$ can only be embedded at  $k=1$ or $k=3$.

\paragraph{Examples.}
We list selected examples in this family below. 
\begin{center}
{
\begin{tabular}{cccccc}
\toprule
Group & $m$ & $(k,d)$ & Fermion content & $(k_3,k_2,k_Y)$ & $\sin^2\theta_W(M_U)$ \\
\midrule
$SU(12)$ & 1 & $(1,1)$ & $6\,\rep{66}+3\,\overline{\rep{78}}$ & $(3,1,1)$ & $1/2$ \\
$SU(20)$ & 2 & $(1,3)$ & $3\,\rep{190}+2\,\overline{\rep{210}}$ & $(5,1,7/3)$ & $3/10$ \\
$SU(28)$ & 3 & $(1,3)$ & $4\,\rep{378}+3\,\overline{\rep{406}}$ & $(7,1,11/3)$ & $3/14$ \\
$SU(28)$ & 3 & $(3,1)$ & $4\,\rep{378}+3\,\overline{\rep{406}}$ & $(7,3,5/3)$ & $9/14$ \\
\bottomrule
\end{tabular}}
\end{center}
For $SU(28)$, we have two inequivalent embeddings, corresponding to $(k,d)=(1,3)$ and $(3,1)$,
respectively. These are explicit anomaly-free fermion constructions
with the required SM chirality.

\paragraph{A 3-index family with color sextets.}
Beyond the two-index family, there are new embeddings that use higher UV representations. We present an example of such a family (but we do not have a proof that this is the only possibility). The matter content of these theories is very large and it will probably prove challenging to build realistic models out of these embeddings.

The family of GUT groups is $SU(n)$ with $n=12+20s$, $s=0,1,\ldots$, and defining
branching $V_s=\overline V_{12}+sB_{20}$, where $V_{12}$ is the
$m=k=d=1$ case of Eq.~\eqref{eq:series-defining}. Here 
$B_{20}=\operatorname{Sym}^2(C+\sigma)
+\overline{\operatorname{Sym}^2(C+\sigma)}$ is a charged real representation
containing a color sextet--antisextet pair. Its indices and predictions,
under the same PQ and matching assumptions as above, are
\begin{equation}
 (k_3,k_2,k_Y)=(3+12s,1,1+8s),\qquad
 \frac EN=\frac23,\qquad
 \sin^2\theta_W(M_U)=\frac{1}{2+8s}.
 \label{eq:sextet-series-indices}
\end{equation}
Writing $A_j=\wedge^j\rep n$, $S_j=\operatorname{Sym}^j\rep n$,
and define the irreducible mixed tensors by
$S_2\otimes\overline{\rep n}=M_S\oplus\rep n$ and
$A_2\otimes\overline{\rep n}=M_A\oplus\rep n$.
An explicit left-handed fermion spectrum is
\begin{equation}
 \mathcal F_s=3\left[S_2+2\overline A_2
 +s\left(S_3+A_3+\overline M_S+\overline M_A\right)\right].
 \label{eq:sextet-series-fermions}
\end{equation}
It obeys $\mathcal A_G(\mathcal F_s)=3(12-n+20s)=0$ and the exact
character identity
$\Delta(\mathcal F_s\!\downarrow_{\rm SM})=3\Delta\mathcal F_{\rm SM}$.
The remaining complex representations pair, and pseudoreal spectators
have even multiplicity, so all spectators admit SM-preserving masses.
The first new member is $SU(32)$, with $\sin^2\theta_W=1/10$;
the sequence continues through $SU(52),SU(72),\ldots$.
These are not asymptotically free for any $s>0$. We leave three-index families for future exploration.

\section{A GUT bootstrap}\label{app:gut-bootstrap}
The search for a consistent embedding of the SM into a unified gauge group is a classification task that is subject to several mathematical constraints. In this respect, it resembles a bootstrap program, where we can use these general constraints to explore the GUT landscape, and islands evading these constraints point to interesting possible new structures. Indeed, the $SU(12)$ case in the text is exactly such an example. In this appendix we derive the relevant constraints on embeddings that contain the SM, and scan over embeddings to find viable GUT models.

\subsection{Assumptions and the odd-trace constraint}
The criteria for the search are the following: we seek a compact, connected, simple gauge
group $G$ to contain the Standard Model (SM) and admit an anomaly-free,
finite Weyl-fermion spectrum $\mathcal F$ with exactly three SM generations
and no additional chiral matter. The spectators must admit
SM-preserving fermion bilinear masses, and hypercharge must generate a
closed circle. For the axion prediction we assume a Peccei--Quinn symmetry
commuting with $G$, with a nonzero mixed gauge anomaly.

Complex spectators are paired, real spectators admit Majorana masses,
and pseudoreal spectators must occur with even multiplicity. The odd
traces for these representation cancel. For an SM Cartan element $X$, the UV spectrum must therefore
satisfy
\begin{equation}
 \Tr_{\mathcal F}X^r=3s_r(X),\qquad
 s_r(X)=\Tr_{\mathcal F_{\rm SM}}X^r,\qquad r=1,3,5,\ldots,
 \label{eq:bootstrap-odd-traces}
\end{equation}
where $\mathcal F_{\rm SM}$ is one SM generation. In particular,
$s_1=s_3=0$, but $s_5\ne0$.

The fifth trace already gives a strong constraint on the embedding. The
nonzero fifth SM trace requires a degree-five invariant of the UV
algebra. The invariant-degree list~\cite{invariants}, together with the SM rank, leaves
only $SU(n)$, $SO(10)$ and $E_6$.
For the latter two, the degree-seven invariant is the degree-five
invariant times the unique quadratic form. Matching the SM fifth and
seventh traces therefore fixes the relative color, weak and hypercharge
normalizations to the canonical ones, giving $E/N=8/3$. For $SU(n)$,
additional independent invariants prevent this immediate conclusion;
the remaining bootstrap is necessary.

\subsection{The $SU(n)$ bootstrap}
In this section we describe a scan over the possible embeddings of the SM in $SU(n)$, and general constraints that such an embedding must satisfy. A key feature of the bootstrap is that the constraints do not require  enumerating the fermion spectrum $\mathcal F$, and therefore can rule out broad classes
of GUTs.

For $G=SU(n)$, an embedding is specified by the branching of the fundamental
representation $V=\rep n$ into color and weak irreducible blocks, with a
given hypercharge. This parametrizes both standard and more exotic embeddings:
\begin{equation}
 \begin{aligned}
 V\downarrow{\rm SM}&=\bigoplus_i c_i(R_i,\ell_i)_{Y_i},\\
 \sum_i c_i\dim(R_i)\ell_i&=n,\qquad
 \sum_i c_i\dim(R_i)\ell_iY_i=0.
 \end{aligned}
 \label{eq:bootstrap-embedding}
\end{equation}
Here $R_i$ is a color representation, $\ell_i$ labels the $SU(2)_W$ representation with dimension $\ell_i$, $c_i$ a positive integer multiplicity, and
$Y_i\in\mathbb Q$ its hypercharge. Both nonabelian factors must act
nontrivially. The two sums impose the dimension budget and traceless
hypercharge. $V$ may be defined only on the simply connected cover,
while $\mathcal F$ should be consistent with the chosen global group.
The embedding indices are defined as:
\begin{equation}
 k_3=\sum_i c_i\ell_i\,2T_3(R_i),\qquad
 k_2=\sum_i\frac{c_i\dim(R_i)\ell_i(\ell_i^2-1)}6,\qquad
 k_Y=2\sum_i c_i\dim(R_i)\ell_iY_i^2.
 \label{eq:bootstrap-indices}
\end{equation}
We use $Q=T_L^3+Y$ and fundamental normalization $\Tr T^a T^b=\delta^{ab}/2$.
Under the stated PQ assumption, the anomalous axion-photon coupling parameter is
\begin{equation}
 \frac{E}{N}=\frac{k_2+k_Y}{k_3}.
 \label{eq:bootstrap-ratio}
\end{equation}

The defining traces are determined by this branching alone:
\begin{equation}
 P_r(X)=\Tr_VX^r
 =\sum_i c_i\,\Tr_{(R_i,\ell_i)_{Y_i}}X^r.
 \label{eq:bootstrap-data-to-traces}
\end{equation}
Every trace in an arbitrary UV representation is an invariant polynomial
of $SU(n)$, and can be expressed in terms of $P_2,\ldots,P_n$.
For example, $\Tr_{\mathcal R}X^5$ is a linear combination of $P_5$ and
$P_2P_3$, the only monomials of total degree five. Summing over any UV
fermion spectrum and using Eq.~\eqref{eq:bootstrap-odd-traces} gives the constrain equation
\begin{equation}
 3s_5=A P_5+B P_2P_3,\qquad A,B\in\mathbb Z.
 \label{eq:bootstrap-fifth-span}
\end{equation}
The integers $A,B$ encode the unknown fermion multiplicities. If no
choice of these constants makes the polynomial identity hold for every
$X$, the embedding is excluded, independently of the UV representation sizes.

We first allow arbitrary rational coefficients to obtain a necessary
test; the additional integer constraints imposed by a physical UV
spectrum are applied below. More generally, one must be able to write 
\begin{equation}
 3s_r=\sum_{\nu_1+\cdots+\nu_k=r}
 c_{\nu_1\cdots\nu_k}\,P_{\nu_1}\cdots P_{\nu_k},
 \qquad c_{\nu_1\cdots\nu_k}\in\mathbb Q,
 \qquad r=5,7,9,\ldots .
 \label{eq:bootstrap-hierarchy}
\end{equation}
The sum runs over distinct products of total degree $r$, with
$2\le\nu_a\le n$ and all product lengths $k$.

\subsection{Additional constraints}

In this subsection we enumerate some of the restrictions on GUT completions of the SM that arise from the constraint equations derived in the previous section. We do not give full derivations of all of our results, instead summarizing the constraints. A detailed derivation of these constraints, together with the code used for the scan, will be presented in a forthcoming publication.

\paragraph{Integer constraints.}
Gauge-anomaly cancellation 
implies that the coefficients in Eq.~\eqref{eq:bootstrap-fifth-span} must be $A=12u$ and $B=10v$, strengthening the degree-five matching condition to:
\begin{equation}
 3s_5=12u\,P_5+10v\,P_2P_3,\qquad u,v\in\mathbb Z.
 \label{eq:bootstrap-integral-fifth}
\end{equation}
These integers parametrize the UV spectrum, so are additional data on the embedding. Matching coefficients gives
further constraints, including that $k_3$ must be odd. There are seventh- and
ninth-degree integral tests that can further restrict the embedding.

\paragraph{Charge realizability.}
For each proposed branching, the trace constraints must admit solutions with rational hypercharges for the specified blocks.

\paragraph{Character compatibility.} The defining branching also determines the full set of eigenvalue 
of $M(g)$. Here $M(g)$ is the $n\times n$ matrix by which an element $g$ of a cover of the SM group, $\widetilde H$, acts on $V$. Write $\Delta\chi_R=\chi_R-\chi_{\overline R}$, where $\chi_R(g)=\Tr_R\,g$ is the character of $R$. For a given fermion spectrum, $\Delta\chi$ is the sum of $\Delta\chi_R$ over its representations, with multiplicity. On the common
cover $\widetilde H$ of the SM where $V$ is defined, any UV
completion must obey
\begin{equation}
 M(g)\sim_{SU(n)} M(h)\quad\Longrightarrow\quad
 \Delta\chi_{\mathcal F_{\rm SM}}(g)=\Delta\chi_{\mathcal F_{\rm SM}}(h),
 \qquad g,h\in\widetilde H.
 \label{eq:bootstrap-character}
\end{equation}
Conjugate matrices have the same character in every UV representation,
so violation of \eqref{eq:bootstrap-character} rejects the embedding independently of the UV representation. Indeed, the character of any $SU(n)$ representation at $g$ depends only on the eigenvalues of $M(g)$, which conjugation by $SU(n)$ transformations does not change. Decomposed under the SM, $\mathcal F$ consists of three generations plus spectators with vanishing $\Delta\chi$ (vectorlike pairs, real and pseudoreal SM representations), so $\Delta\chi_{\mathcal F}=3\,\Delta\chi_{\mathcal F_{\rm SM}}$. Conjugate $M(g)$ and $M(h)$ thus give Eq.~\eqref{eq:bootstrap-character}.
The scan combines these tests with integral information through degree
nine, enumerating nonabelian blocks at fixed $n$ while keeping charges
symbolic. The tests are necessary but not sufficient to establish an anomaly free UV spectrum without chiral exotics. Embeddings which pass the tests highlight interesting candidates, leading to families of viable models, as we will see below.

\paragraph{Explicit checks used in the scan.}
At zero hypercharge, the integer constraint can be evaluated using
three coefficients computed from the defining branching:
\begin{equation}
 \begin{aligned}
 C_3&=\sum_i c_i\ell_i\mathcal A_3(R_i),\qquad
 C_{3,2}=\sum_i c_i\mathcal A_3(R_i)\frac{\ell_i(\ell_i^2-1)}6,\\
 P_5(Z)&=\frac56 C_5\,\Tr_{\rep3}Z^2\Tr_{\rep3}Z^3,
 \end{aligned}
 \label{eq:bootstrap-filter-coefficients}
\end{equation}
where $\mathcal A_3(\rep3)=1$ and $Z$ is a pure color Cartan element.
For the classified triplet embeddings, $C_3$ equals the imbalance $d$.
The candidate is rejected unless $u,v$ simultaneously satisfy 
\begin{equation}
 uC_5+v k_3C_3=0,\qquad 12uC_{3,2}+v k_2C_3=3.
 \label{eq:bootstrap-integer-filter}
\end{equation}
In particular, $k_2,C_3$ are odd, and $C_3\equiv k_3\pmod2$ forces
$k_3$ odd. For the $SU(12)$ branching in Eq.~\eqref{eq:fund_decomposition},
\begin{equation}
 (n,k_3,k_2,k_Y;C_3,C_{3,2},C_5)=(12,3,1,1;1,0,1),\qquad
 u+3v=0,\quad v=3.
 \label{eq:bootstrap-su12-filter}
\end{equation}
Thus $u=-9$. The spectrum $6\rep{66}+3\overline{\rep{78}}$ realizes
these coefficients. Its full branching in Sec.~\ref{app:branching}
confirms a possible UV construction.

\subsection{Results and coverage}

\paragraph{Theorem (two-index classification).}
For every $n\ge5$, the admissible $SU(n)$ embeddings and net chiral
spectra with UV fermions carrying at most two indices in total,
fundamental or antifundamental, consist of the two canonical and three noncanonical
branches listed in Table~\ref{tab:two-index-families} and proved in Sec.~\ref{app:two-index-proof}.
These families of $SU(n_0)$ admit simple extensions to $SU(n_0 +r)$ with $r$ additional singlets in the defining block. Together, this is an exhaustive classification. All five branches have exactly three chiral
SM generations and vectorlike spectators.
The canonical branches (families I \& II) have $E/N=8/3$, with $k_3=1$ for $n\ge5$
or $k_3=3$ for $n\ge15$~\cite{fonseca}.
With $K=k_3$ odd, the three noncanonical branches project to
\begin{equation}
 n=4K+r,\quad K\ge3,\quad r\ge0,\qquad E/N=2/3.
 \label{eq:bootstrap-two-index-region}
\end{equation}
Here $r$ counts neutral singlets added to the charged core.

\begin{figure}[tbp]
\centering
\resizebox{0.9\linewidth}{!}{\input{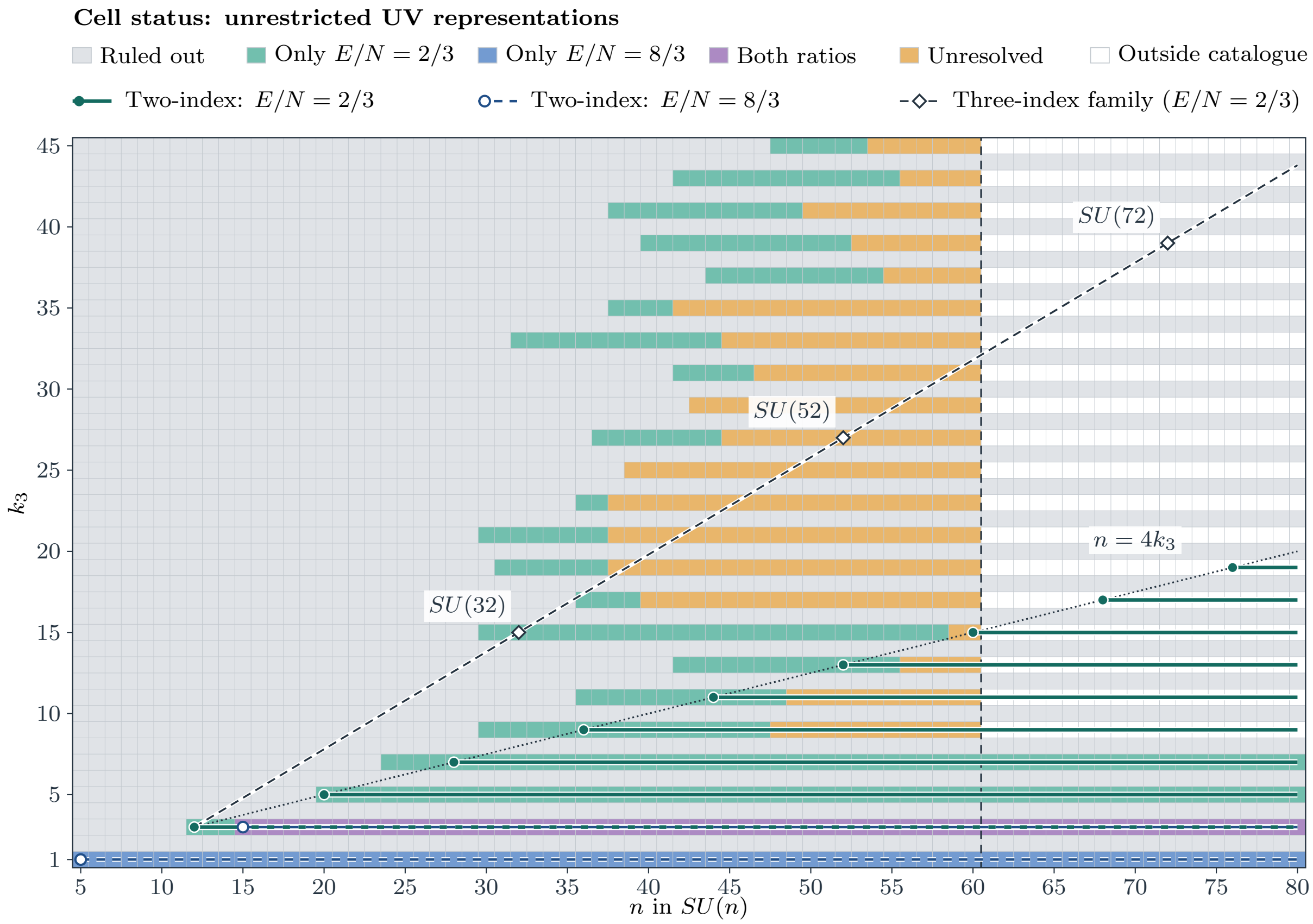}}
\caption{\small Bootstrap coverage and explicit families for $SU(n)$.
Each cell labels $(n,k_3)$, with the background colors showing the status: ruled out, only $E/N=2/3$, only $8/3$, both
ratios among surviving candidates, or unresolved. Ratio colors are
necessary conditions. Solid teal and dashed blue horizontal lines show the
exhaustive two-index constructions, starting at $n=4k_3$. The canonical GUTs are $(n,k_3)=(5,1)$ and $(15,3)$.
 Diamonds mark the known three-index family
$(n,k_3)=(12+20s,3+12s)$, $s\ge1$, admitting a three-index completion
but no two-index completion. The vertical dashed line marks the $n\le60$ catalogue boundary.
Beyond it, only all-rank results are colored; white cells are outside
coverage. We note that due to the matching condition in Eq.~\eqref{eq:matching_condition}, models with $k_3 \gtrsim \alpha_{3}^{-1}(M_U)$ correspond to strongly coupled theories $\alpha_{\rm GUT}>1$. All theories here contain three chiral SM families and vectorlike matter.}
\label{fig:bootstrap}
\end{figure}

The family of Eqs.~\eqref{eq:sextet-series-indices}--\eqref{eq:sextet-series-fermions}
instead gives
\begin{equation}
 (n,k_3)=(12+20s,\,3+12s),\qquad s\ge1.
 \label{eq:bootstrap-three-index-family}
\end{equation}
The displayed spectrum uses three-index fermions, while
Eq.~\eqref{eq:bootstrap-two-index-region} and the canonical branches
exclude any two-index completion. These are constructions, not an exhaustive
three-index classification.

\begin{center}
\small
\begin{tabular}{lll}
\toprule
UV tensors & Domain & Established result \\
\midrule
Up to two indices & All $n$ & Embeddings and net chiral spectra classified \\
Up to three indices & $n=12+20s$, $s\ge1$ & New family; two-index completion excluded \\
Unrestricted & $n\le37$ or $k_3\le7$ & $E/N\in\{2/3,8/3\}$ \\
Unrestricted & Triplet-only colored blocks, all $n$ & $E/N\in\{2/3,8/3\}$ \\
\bottomrule
\end{tabular}
\end{center}
In the last row, colored defining blocks are weak-singlet triplets or
antitriplets; colorless blocks are arbitrary. The unrestricted finite-rank
and low-index results include certified finite calculations. They fix
the ratio if a completion exists, rather than classify all spectra.
The unrestricted all-rank two-value theorem remains open.

Beyond the two-index cutoff, UV-chiral sectors can be SM-nonchiral and
anomaly-free~\cite{fonseca}, so classifying embeddings need not classify
all UV spectra. The plot therefore distinguishes a complete low-degree
classification from particular higher-degree constructions.

\section{Classification with two-index fermions}\label{app:two-index-proof}
In this section we outline the proof for the two-index classification given in Table~\ref{tab:two-index-families} for every $SU(n)$, $n\ge5$ (up to trivial extensions such as adding singlets to the branching of $V$).
The two-index restriction means that the UV fermions may occupy the fundamental $V=\rep n$,
its antisymmetric square $A_2=\wedge^2V$ and symmetric square
$S_2=\operatorname{Sym}^2V$, their conjugates, and the real
singlet and adjoint, with arbitrary multiplicities. We require
cancellation of the UV gauge anomaly and exactly three SM generations
of net chirality, with all other fermions admitting SM-preserving vectorlike
masses. The branching of $V$, which specifies the embedding, is
initially unrestricted.

The argument first bounds the weak content using a single group
element and a weighted anomaly. Ordinary spin addition then excludes defining blocks carrying
both color and weak charge. Finally, nonnegative charge multiplicities
fix hypercharge. The result is the two canonical and three noncanonical
families of Sec.~\ref{app:su8m4-family}.

\subsection{Weighted Anomaly Matching}
We show that a form of weighted anomaly matching can put a sharp restriction on the allowed embedding. Up to additional real and vectorlike components the GUT fermion spectrum can be written as
\begin{equation}
    \mathcal{F}
    =
    x V
    +y A_2
    +z S_2,
\end{equation}
where $x,y,z$ are signed integers, with a negative coefficient denoting
the conjugate representation. It will be useful to define $p=y+z$.
The $SU(n)$ anomaly coefficients in these representations are $1,n-4,n+4$, so anomaly cancellation implies
\begin{equation}
 x+(n-4)y+(n+4)z=0,\qquad x=-np-4(z-y).
 \label{eq:two-index-uv-anomaly}
\end{equation}

We can get additional constraints using a weighted anomaly equation. For a $SU(n)$ representation $\mathcal{R}$ branching to SM, 
\begin{equation}
    \mathcal{R}\downarrow SM= \bigoplus c_i(R_i,\ell_i)_{y_i},
\end{equation}
define
\begin{equation}
    \mathcal A_g(\mathcal{R})
    =
    \sum_i \mathcal A_{3}(c_i R_i) \chi_{\ell_i}(g)
\end{equation}
where $\mathcal A_3$ is the $SU(3)^3$ anomaly coefficient and the $SU(2)$ trace $\chi_{\ell}(g)= \Tr_{\ell} g$ for some element $g$. We will choose the weak group element
\begin{equation}
 g=\operatorname{diag}(\omega,\omega^2),\qquad
 \omega=e^{2\pi i/3}.
\end{equation}
Its $\omega$ and $\omega^2$ eigenspaces in $V$ have equal dimension
$k$, since $g$ and $g^{-1}$ are conjugate in $SU(2)$. Thus
\begin{equation}
 P:=\Tr_V g=n-3k.
\end{equation}
Every nontrivial weak multiplet contributes at least one eigenvector
to each of these eigenspaces. 
A defining block with color dimension $d_c$ therefore contributes at
least $d_c$ to $k$ if it is a weak nonsinglet. In particular,
$k$ is a positive integer.

Importantly, $\mathcal A_g$ depends only on the chiral spectrum (with contributions from vectorlike matter vanishing identically), so can be matched with the SM,
\begin{equation}
    \mathcal A_g(\mathcal{F})
    =3 \mathcal A_g(\mathcal F_{\rm SM})
    =-9
\end{equation}
where $\mathcal F_{\rm SM}$ is the fermion content in one generation of SM fermions. The quark doublet contributes $-1$ and the antiquarks $-2$ to the weighted anomaly.

Define $d = \mathcal A_g(V)$.
With $\mathcal A_3(\mathbf{3})=1$, the color anomaly coefficients
and the weak characters at the chosen $g$ are integers,
so $d\in\mathbb Z$.
To extract the weighted anomaly for the two-index representations,
consider $f(t)=\Tr_V(e^{tX_c}g)$, with $X_c$ a color
generator normalized so $\Tr_{\rep3}X_c^3=1$. Then $f'(0)=0$,
$f(0)=P$, and $f'''(0)=d$.
Using the identity,
\begin{align}
\Tr_{A_2}(e^{t X_c}g)
=
\frac12\left[f(t)^2 -\Tr_V(e^{2tX_c}g^2)\right]
\\
\Tr_{S_2}(e^{t X_c}g)
=
\frac12\left[f(t)^2 +\Tr_V(e^{2tX_c}g^2)\right]
\end{align}
and differentiating the expression three times,
we obtain ($g$ and $g^2$ are related by conjugation),
\begin{equation}
 \mathcal A_g(A_2)=(P-4)d,\qquad
 \mathcal A_g(S_2)=(P+4)d.
 \label{eq:two-index-weighted-squares}
\end{equation}
These weighted anomaly formulae are the usual anomaly formulae with $n$ replaced by $P$. Matching the weighted anomaly with the SM gives
\begin{equation}
 [x+(P-4)y+(P+4)z]d=-3pkd = -9,
 \qquad \boxed{pkd=3}.
 \label{eq:two-index-order-three-budget}
\end{equation}
Consequently $k=1$ or $3$, independently of $n$.

\subsection{Restricting the defining blocks}

Given the bound on $k$, we can derive constraints on the branching of the fundamental representation.

\paragraph{No simultaneous color and weak charge.}
In the branching of the fundamental representation such a block contributes at least three to $k$, therefore the bound permits just one
copy with color $R=\rep3$ or $\overline{\rep3}$ with all other defining
blocks being weak singlets. Such a branching cannot reproduce the SM chiral spectrum. 
More explicitly, let the weak representation of this block be labelled by its spin, $j>0$. Then its square in the cross-product contained in $A_2$ or $S_2$ has a rep $\tilde R$  has weak spin $2j \geq 1$ and colour charge, while all other representations have weak spin $0$ or $j$. Thus, there is an exotic chiral fermion in the $\tilde R$ rep left over if representations in the branching of $V$ carry both color and weak charges.

\paragraph{Match quark doublets and antiquarks.}
With hypercharge omitted for the time being, write $V=U\oplus W$, where $U$ contains
the weak singlets and $W$ the colorless weak nonsinglets.
All colored weak nonsinglets in the branching of the two-index representations arise from the cross product
$U\otimes W$. It appears once in each two-index representation,
so its signed coefficient is $p=y+z$. Since $\overline W=W$
when hypercharge is omitted, matching net chirality three quark doublets requires
\[
 p\,(U-\overline U)\otimes W
 =3(\rep3-\overline{\rep3})\otimes\rep2.
\]
The right-hand side contains only weak doublets, so $W$ must
contain only doublets. Each contributes one to the eigenspace dimension
$k$. Since all colored blocks are weak singlets, $d=\mathcal A_g(V)$
is now the ordinary color anomaly. Matching the color representations and
multiplicities therefore gives
\begin{equation}
 W=k\,\rep2,\qquad
 U-\overline U=d(\rep3-\overline{\rep3}),\qquad pkd=3.
 \label{eq:two-index-generation-factors}
\end{equation}
Choosing the convention $d>0$,
$(k,d)=(1,1),(1,3),(3,1)$ and $p=3/(kd)$. We can thus conclude from eq.~\eqref{eq:bootstrap-indices} that $k$ gives the weak embedding level, $k = k_2$.

We can also fix the unsigned color content as well.
Define $K=(y-z)/p$, which is well-defined since $pkd=3$.
Some algebra with symmetric and antisymmetric squares gives,
\begin{equation}
 d(\rep3-\overline{\rep3})\otimes
 \left[x\,\rep1+\frac p2(U+\overline U)
       -\frac {pK}2(\rep3+\overline{\rep3}+2\,\rep1)\right]
 =-6(\rep3-\overline{\rep3}).
\end{equation}
Using Eqs.~\eqref{eq:two-index-uv-anomaly} and
\eqref{eq:two-index-generation-factors}, this simplifies to,
\begin{equation} \label{eq:UplusUbar}
 U+\overline U=K(\rep3+\overline{\rep3})
               +(2n-6K-4k)\rep1.
\end{equation}
Thus, every higher color
irrep is also excluded. We can restrict the branching to,
\begin{equation}
 \begin{gathered}
 V=a(\rep3,\rep1)+b(\overline{\rep3},\rep1)
      +k(\rep1,\rep2)+s(\rep1,\rep1),\\
 a-b=d,\qquad a+b=K.
 \end{gathered}
 \label{eq:two-index-nonabelian-content}
\end{equation}
The integers $p,K$ are odd, since $pkd=3$ and $K=d+2b$. By Eq.~\eqref{eq:UplusUbar} the only colored blocks in $V$ are the $a$ triplets and $b$ antitriplets of Eq.~\eqref{eq:two-index-nonabelian-content}, all of them weak singlets, and since each contributes $\ell_i\,2T_3(R_i)=1$ to the color index in Eq.~\eqref{eq:bootstrap-indices}, we have $k_3=a+b=K$.

\subsection{Hypercharge and the surviving families}

We now assign hypercharges to the defining blocks already found.
The strategy is to use sextets that appear in the product representations and require they do not appear in the chiral spectrum, and to use the color and weak singlets.

\paragraph{Cancel the color sextets.}
Two defining (anti)triplet blocks can produce a (anti)sextet in a two-index
representation.
In the fermion spectrum, all sextets must be paired with antisextets of opposite hypercharge. Consider the defining branching of the fundamental representation. Each of the triplets and antitriplets can carry different hypercharges. Cancellation of chiral sextets forces all triplets to carry the same hypercharge $h$ and antitriplets to carry $-h$.

Explicitly, let $a_Y$ be the multiplicity for each triplet with hypercharge $Y$, so that $\sum a_Y = a$. Similarly, define $b_Y$ as the multiplicity of antitriplets with hypercharge $-Y$. Choose $h$ to be the highest hypercharge for which $a_Y \neq b_Y$. Such an $h$ exists since $a\neq b$. If there is a triplet with hypercharge $H > h$ in the spectrum, by definition of $h$, $a_H = b_H$. Products of these hypercharged triplets/antitriplets with the blocks at $h$
then leave unequal numbers of sextets at $H+h$ and antisextets at
$-(H+h)$, which cannot be the case as we do not allow chiral exotica. Therefore, $h$ is the highest hypercharge for triplets in the defining branching.

This implies that sextets at $Y = 2h$ and antisextets at $Y = -2h$ arise from the triplets at $h$ and
antitriplets at $-h$. 
Counting their pairs in the two-index representations gives
\begin{equation}
 \begin{aligned}
 0&=y\left[\binom{a_h}{2}-\binom{b_h}{2}\right]
    +z\left[\binom{a_h+1}{2}-\binom{b_h+1}{2}\right]\\
  &=\frac p2(a_h-b_h)(a_h+b_h-K),
 \end{aligned}
\end{equation}
where $z-y=-pK$. This means $a_h+b_h=K=a+b$, the total number of colored
defining blocks. Every defining triplet therefore has hypercharge $h$, and every defining antitriplet has charge $-h$.

The SM quark doublet comes from a triplet--doublet product with total hypercharge $1/6$, so the defining doublet that pairs with a triplet of charge $h$ has charge $c=\frac16-h$. The only hypercharges for the defining weak doublets are $\pm c$; any other hypercharge will leave behind chiral exotica. For $c\ne0$, their multiplicities are fixed by asking for 3 quark doublets and cancelling the unwanted $h-c$ representations.
\begin{equation}
 W=\frac{ka}{K}(\rep1,\rep2)_c
   +\frac{kb}{K}(\rep1,\rep2)_{-c},
 \qquad c=\frac16-h.
 \label{eq:two-index-weak-charges}
\end{equation}
If $c=0$, the two contributions combine into $k$ hypercharge neutral doublets. 

\paragraph{Use nonnegative singlet multiplicities.}
The final constraint comes from the hypercharges of the color and weak singlets. 
Let $Y_1,\ldots,Y_s$ be the hypercharges of the defining
color-and-weak singlets, and denote such a singlet by $\rep1_Y$.
Matching the antiquark charges $-2/3$ and $1/3$, and adding the
charge-conjugate equation, gives
\begin{equation}
 \begin{aligned}
 \sum_{i=1}^{s}(\rep1_{Y_i}+\rep1_{-Y_i})
 &=2(n-4K)\rep1_0+K(\rep1_{3h}+\rep1_{-3h})-k(\rep1_{h-2/3}+\rep1_{-h+2/3}
          +\rep1_{h+1/3}+\rep1_{-h-1/3}).
 \end{aligned}
 \label{eq:two-index-positive-charges}
\end{equation}
The left side counts actual singlets and their conjugates, so all
multiplicities are nonnegative. Each negative coefficient on the
right must be cancelled by a positive coefficient at the same
hypercharge. The only possible positive contributions are at
$0$ or $\pm3h$.
Consequently
\begin{equation}
 h-\frac23,\ h+\frac13\in\{0,\pm3h\}
 \quad\Longrightarrow\quad h=-\frac13\ \hbox{or}\ \frac16.
 \label{eq:two-index-charge-values}
\end{equation}

Summing squared charges in Eq.~\eqref{eq:two-index-positive-charges}
fixes the singlet contribution to the hypercharge index. Including
the colored blocks and weak doublets gives
\begin{equation}
 \begin{aligned}
 \sum_{i=1}^{s}Y_i^2
 &=9Kh^2-k\left[\left(h-\frac23\right)^2
                    +\left(h+\frac13\right)^2\right],\\
 k_Y&=2\left(3Kh^2+2kc^2+\sum_{i=1}^{s}Y_i^2\right)=24Kh^2-k.
 \end{aligned}
\end{equation}
Since $k_3=K$ and $k_2=k$, a commuting PQ symmetry with nonzero mixed anomaly therefore has
\begin{equation}
 \boxed{\frac EN=\frac{k+k_Y}{K}=24h^2
                   \in\left\{\frac83,\frac23\right\}.}
\end{equation}

\subsection{Five families}
The constraints above leave behind five families of GUTs, each of which can be consistently realized at the level of fermion content. In the branchings,
$\rep1_0=(\rep1,\rep1)_0$. The anomaly condition fixes the UV multiplicities in Table~\ref{tab:two-index-families} from the values of $p$ and $K$
already determined above. The table lists the gauge group, defining branching $V\downarrow SM$, and
UV Weyl fermion spectrum $\mathcal F_0$ together for each family.
It displays the cores without neutral singlets in $V$; their common
neutral-singlet extension is given below. The only free integer in
the list is $m$ for families III--V, with $K=k_3=2m+1$.
Of course, the matter content is defined up to vectorlike and real representations.

\emph{Families I and II: $h=-1/3$, $E/N=8/3$.}
The doublet charges in
Eq.~\eqref{eq:two-index-weak-charges} are $\pm1/2$. Their squares give
net weak-triplet multiplicity $3(k-K)/(2K)$ at charge $+1$.
Its absence requires $k=K$. Equation~\eqref{eq:two-index-positive-charges}
then leaves only $n-5K$ neutral singlets. The conditions $pkd=3$ and
$d\le K$ give $(K,k,d)=(1,1,1)$ or $(3,3,1)$.
These are families I and II in Table~\ref{tab:two-index-families}.

\emph{Families III--V: $h=1/6$, $E/N=2/3$.}
Equation~\eqref{eq:two-index-positive-charges} fixes the
total number of singlets at charges $\pm1/2$, and traceless
hypercharge fixes their difference. Thus
\begin{equation}
 \bigoplus_{i=1}^{s}\rep1_{Y_i}
 =(a-k)\rep1_{-1/2}+(b-k)\rep1_{1/2}+(n-4K)\rep1_0.
\end{equation}
Nonnegative multiplicities require $a,b\ge k$ and $n\ge4K$.
Together with Eq.~\eqref{eq:two-index-generation-factors}, these give
$(k,d)=(1,1),(1,3),(3,1)$, respectively. Substituting their fixed
multiplicities and setting $K=2m+1$ gives families III--V in
Table~\ref{tab:two-index-families}, all with group $SU(8m+4)$.
All displayed multiplicities are nonnegative. Families III--V start
at $SU(12)$, $SU(20)$ and $SU(28)$, respectively.

\paragraph{Extensions.}
Every listed $SU(n_0)$ can be trivially to $SU(n_0+r)$ for any integer
$r\ge0$: add $r\rep1_0$ to the defining branching, interpret
$V,A_2,S_2$ in the spectrum as representations of the larger group,
and add $3r\overline V$ in families I and III, or $r\overline V$ in
families II, IV and V. This gives $SU(5+r)$, $SU(15+r)$ and
$SU(8m+4+r)$, respectively. The extension preserves UV anomaly
cancellation and adds only vectorlike or neutral SM fermions, so
the net SM chirality and $E/N$ are unchanged.
The general allowed UV spectrum can also contain
arbitrary conjugate pairs of the allowed UV representations and
arbitrary singlets and adjoints to $\mathcal F_0$.

\paragraph{Chirality, masses, and scope}
Write $\Delta R=R-\overline R$ for net chirality.
For the canonical families,
$\Delta\mathcal F_0=3\Delta(\wedge^2F_5+\overline F_5)
=3\Delta\mathcal F_{\rm SM}$.
For the noncanonical ones, expanding the two-index representations gives
\begin{equation}
 \begin{aligned}
 \Delta\mathcal F_0
 &=pkd\,\Delta\Bigl[(\rep3,\rep2)_{1/6}
       +(\overline{\rep3},\rep1)_{-2/3}
       +(\overline{\rep3},\rep1)_{1/3}
       +(\rep1,\rep2)_{-1/2}+(\rep1,\rep1)_1\Bigr]\\
 &=3\Delta\mathcal F_{\rm SM}.
 \end{aligned}
\end{equation}
The bracket is precisely one SM generation.

Complex spectators occur in conjugate pairs. All self-conjugate spectators are
real: half-integer weak-spin states have nonzero hypercharge or
complex color. Adding $r$ neutral defining singlets 
to $V_0$ adds $yr(V_0+\overline V_0)$ and neutral singlets to the
positive fermion spectrum. Thus pseudoreal spectators occur in even
multiplicities as well. All spectators admit SM-preserving mass
bilinears. Arbitrary UV-vectorlike pairs and real UV representations
within the cutoff can of course also be added, since they already admit UV-invariant masses.

\end{document}